\documentclass[10pt,conference]{IEEEtran}

\usepackage{times,graphics,amssymb,amsmath,bm,cite}
\usepackage[dvips]{graphicx}
\usepackage[english]{babel}
\usepackage{latexcad}
\usepackage{epsfig,psfig,latexsym}

\newcommand{\eeq}{\end{equation}}

\newcommand{\br}{\mbox{\boldmath $r$}}

\newcommand{\bs}{\mbox{\boldmath $s$}}

\newcommand{\bP}{\mbox{\boldmath $P$}}

\newcommand{\bH}{\mbox{\boldmath $H$}}

\newcommand{\bS}{\mbox{\boldmath $S$}}

\newcommand{\bn}{\mbox{\boldmath $n$}}

\newcommand{\bd}{\mbox{\boldmath $d$}}

\newcommand{\bI}{\mbox{\boldmath $I$}}

\newcommand{\ds}{\displaystyle}

\newcommand{\beq}{\begin{equation}}

\newcommand{\gammabar}{\bar{\gamma}}

\begin{document}

\title{Power Control Algorithms for
CDMA Networks Based on Large System Analysis}

\author{
\authorblockN{Stefano Buzzi}
\authorblockA{Universit\`{a} degli Studi di Cassino \\
03043 Cassino (FR) - ITALY \\
buzzi@unicas.it}
\and
\authorblockN{H. Vincent Poor}
\authorblockA{School of Engineering and Applied Science \\
Princeton University,
Princeton, NJ, 08544, USA \\
poor@princeton.edu}
}
%

\maketitle

\begin{abstract}
Power control is a fundamental task accomplished in any wireless cellular network; its aim is to set the transmit power of any mobile terminal, so that each user is able to achieve its own target SINR. While conventional power control algorithms require knowledge of a number of parameters of the signal of interest and of the multiaccess interference, in this paper it is shown that in a large CDMA system much of this information can be dispensed with, and effective distributed power control algorithms may be implemented with very little information on the user of interest. An uplink CDMA system subject to flat fading is considered with a focus on the cases in which a linear MMSE receiver and a non-linear MMSE serial interference cancellation receiver are adopted; for the latter case new formulas are also given for the system SINR in the large system asymptote. Experimental results show an excellent agreement between the performance and the power profile of the proposed distributed algorithms and that of conventional ones that require much greater prior knowledge.
\end{abstract}

\section{Introduction} \label{sec:introduction}

In multiuser wireless communication systems, mobile users vary their transmit powers so as to counteract channel gain variations and achieve their target Signal-to-Interference plus Noise Ratios (SINRs); this task is usually referred to as power control \cite{yates}. In code division multiple access (CDMA) wireless systems, power control, possibly coupled with the use of multiuser detectors, is thus used in the uplink to combat the near-far effect, to manage interference levels, and to minimize the overall power radiated by the system. A considerable amount of work has been done on power control algorithms for cellular networks, see, e.g. \cite{yates,nara2,stoc}, to cite a few; these algorithms, based on measurements taken at both the mobile station (MS) and at the base station (BS), give as output the transmit power for each terminal. In power control procedures, estimates of several parameters such as the channel gains and the SINR for each user are usually needed, or, alternatively, recursive algorithms are adopted (see, e.g. \cite{stoc}), which however are affected by slow convergence speed and excess steady-state error.

In recent years a new mathematical tool has emerged in the analysis of CDMA systems, i.e. the so-called large system analysis, first introduced in \cite{tse}. In short,
\cite{tse} has revealed that,
in a CDMA system whose processing gain and number of users both increase without bound with their ratio fixed,
and with randomly chosen unit-norm spreading codes, the SINR at the output of a linear minimum mean square error (MMSE)
receiver converges in probability to a non-random constant.
In particular, denoting by $K$ the number of active users, by $N$ the system processing gain, by $\sigma^2$ the additive thermal noise power spectral density (PSD) level, and by $E_P [ \cdot]$ the expectation with respect to the limiting empirical distribution $F$ of the received powers of the interferers, the SINR of the MMSE receiver for the $k$-th user, say $\gamma_k$, converges in probability  as $K, N \rightarrow \infty$, $K/N=\alpha=\mbox{constant}$, to $\gamma_k^*$ the unique solution of the equation
\beq
\gamma_k^*= \ds \frac{P_k}{\sigma^2 + \alpha E_P \left[
\frac{P P_k}{P_k +P \gamma_k^*}
\right]} \; ,
\label{eq:HT}
\eeq
with $P_k$ the received power for the $k$-th user. Interestingly, the limiting SINR depends only on the limiting empirical distribution of the received powers of the interferers,  the load $\alpha$, the thermal noise level and the transmit power of the user of interest, while being independent of the actual realization of the received powers of the interferers and of the spreading codes of the active users. Large system analysis is now a well-established mathematical tool  for the design and analysis of communication systems (see, e.g., \cite{evans1,mimo}, to cite a few).

In this paper, we show how large system analysis can be used to design distributed power control algorithms that need very little prior information (i.e. the channel gain for the user of interest) to be implemented,  for both the cases in which linear MMSE detection and non-linear interference cancellation MMSE detection are used at the receiver. The contributions of this paper can thus be summarized as follows.
\begin{itemize}
\item[-]
Assuming that   linear MMSE detection is used at the receiver, we propose a new distributed power control algorithm requiring little prior information and based on large system analysis.
\item[-]
Extending the approach in \cite{tse}, we give an expression for the limiting SINR in the case in which non-linear serial interference cancellation MMSE (SIC/MMSE) detection is adopted at the receiver.
\item[-]
Assuming that SIC/MMSE detection is used at the receiver, we propose a new distributed power control algorithm based on large system  approximations.
\item[-]
We show how the proposed algorithms can be included in utility-maximizing non-cooperative games in order to achieve energy-efficiency in wireless data networks \cite{meshkati}.
\end{itemize}
Our numerical results will show that the proposed algorithms are effective and are able to approach very closely the power profile predicted by much more complex and information-demanding power control algorithms.


\section{System model and problem statement}
Consider the uplink of a $K$-user synchronous, single-cell, direct-sequence code division multiple access (DS/CDMA) system with processing gain $N$ and subject to flat fading. After chip-matched filtering and sampling at the chip-rate, the $N$-dimensional received data vector, say $\br$, corresponding to one symbol interval, can be written as
\beq
\br=\ds \sum_{k=1}^{K}\sqrt{p_k} h_k b_k \bs_k + \bn \; ,
\label{eq:r}
\eeq
wherein $p_k$ is the transmit power of the $k$-th user\footnote{To simplify subsequent notation, we assume that the transmitted power $p_k$ subsumes also the gain of the transmit and receive antennas.}, $b_k\in \{-1,1\}$ is the information symbol of the $k$-th user, and $h_k$ is the real\footnote{For simplicity, we assume a real channel model; however, generalization to practical channels with I and Q components is straightforward.} channel gain between the $k$-th user transmitter and the base station; the actual value of $h_k$ depends on both the distance of the $k$-th user's mobile from the base station and the channel fading fluctuations. The $N$-dimensional vector $\bs_k$ is the spreading code of the $k$-th user; we assume that the entries of $\bs_k$ are binary-valued and that $\bs_k^T \bs_k=\|\bs_k\|^2=1$, with $(\cdot)^T$ denoting transpose. Finally, $\bn$ is the thermal noise vector, which we assume to be a zero-mean white Gaussian random process with covariance matrix $\sigma^2 \bI_N$, with $\bI_N$ the identity matrix of order $N$.

Given the system model (\ref{eq:r}), a number of strategies are available to detect the data symbols $b_1, \ldots, b_K$; in the following we briefly review the linear MMSE receiver and the non-linear SIC/MMSE receiver.

\subsection{Linear MMSE detection}
Consider the case in which a linear receiver is used to detect the data symbol $b_k$, according, i.e., to the decision rule
$
\widehat{b}_k=\mbox{sign}\left[\bd_k^T \br\right] \; ,
$
with $\widehat{b}_k$ the estimate of $b_k$ and $\bd_k$ the $N$-dimensional vector representing the receive filter for the user $k$. Then, it is easily seen that the
linear MMSE receiver is the one corresponding to the choice
$
\bd_k=\sqrt{p_k} h_k \left(\bS \bH \bP \bH^T  \bS^T + \sigma^2 \bI_N\right)^{-1} \bs_k \; ,
$
wherein $\bS=[ \bs_1, \ldots, \bs_K]$ is the $N\times K$-dimensional spreading-code matrix, and $\bP$ and $\bH$ are $K \times K$-dimensional diagonal matrices, whose diagonals are $[p_1, \ldots, p_K]$ and $[h_1, \ldots, h_K]$, respectively. For linear detectors it is also meaningful to define the output SINR, which, for the $k$-th user is written as
\beq
\gamma_k=\ds \frac{p_k h_k^2 (\bd_k^T \bs_k)^2}{\sigma^2\|\bd_k\|^2 + \ds \sum_{i \neq k} p_i h_i^2
(\bd_k^T \bs_i)^2} \; .
\label{eq:gamma}
\eeq

\subsection{Non-linear SIC/MMSE detection}
Consider now the case in which  non-linear decision feedback detection is used at the receiver.  We assume that the users are indexed according to a non-increasing sorting of their channel gains, i.e. we assume that $h_1 > h_2 > \ldots, h_K$. We consider a serial interference cancellation (SIC) receiver wherein detection of the symbol from the $k$-th user is made according to the rule
$
\widehat{b}_k=\mbox{sign}\left[\bd_k^T \br_k  \right] \; ,
$ wherein $\br_k=\br - \sum_{j<k} \sqrt{p_j} h_j \widehat{b}_j \bs_j$.

Otherwise stated, when detecting a certain symbol, the contributions from the data symbols that have already been detected are subtracted from the received data. The output SINR for user $k$, under the assumption of correctness of past decisions, is now written as
\beq
\gamma_k=\ds \frac{p_k h_k^2 (\bd_k^T \bs_k)^2}{\sigma^2\|\bd_k\|^2 + \ds \sum_{j > k} p_j h_j^2
(\bd_k^T \bs_j)^2} \; .
\label{eq:gammaSIC}
\eeq
Upon defining $\bS_k=[\bs_{k}, \ldots, \bs_K]$, $\bP_k=\mbox{diag}(p_{k}, \ldots, p_K)$ and
$\bH_k=\mbox{diag}(h_{k}, \ldots, h_K)$, it is easy to show that SIC/MMSE detection corresponds to the choice
$
\bd_k=\sqrt{p_k} h_k(\bS_k \bH_k \bP_k \bH_k^T \bS_k^T + \sigma^2\bI_N)^{-1} \bs_k \; .
$

\subsection{Problem statement}
Given the data model (\ref{eq:r}), we are interested in the following problem: find the transmit power $p_k \in [0, P_{\max}]$ for each user $k$, so that the SINR $\gamma_k$ equals a given target value $\bar{\gamma}$, with $P_{\max}$ the maximum power that each user in the system is allowed to transmit. Note that this problem finds numerous  applications. As an example, in circuit-switched wireless cellular voice communications, wherein the primary goal is signal intelligibility, the SINR is required to be always above a given intelligibility threshold \cite{yates}. In wireless packet-switched networks, instead, the primary goal may be to maximize the system throughput, or, if battery-life of the mobile terminals is a dominant issue, the system throughput for each unit or energy drained from the battery \cite{nara2,meshkati}. In all cases, however, it can be shown that this goal translates into the requirement that each user's SINR equals at least a certain value.

In the sequel, we show how large system analysis leads to power control algorithms that may be implemented in a distributed fashion and that require knowledge of the channel for the user of interest only.

\section{Power Control for linear MMSE detection}

As an introductory step in our algorithm, we begin by illustrating a simple power control algorithm derived from \cite{tse}.
We have seen that in a large CDMA system the $k$-th user's SINR converges in probability to the solution to Eq. (\ref{eq:HT}). Heuristically, this means that in a large system, and embracing the notation of the previous section, the SINR $\gamma_k$ is deterministic and approximately satisfies
\beq
\gamma_k\approx \ds\frac{h_k^2 p_k}{\sigma^2 +  \frac{1}{N} \sum_{j \neq k} \frac{h_k^2 h_j^2 p_k p_j}{h_k^2 p_k + h_j^2 p_j \gamma_k}}
\label{eq:HTheur}
\eeq
Now, as noted in \cite{tse}, if all the users must achieve the same common target SINR $\bar{\gamma}$, it is reasonable to assume that they are to be received with the same power, i.e. the condition
$
h_1^2 p_1= h_2^2 p_2 = \ldots h_K^2 p_k = P_R\; ,
$
is to be fulfilled. Substituting the above constraint in (\ref{eq:HTheur}) and equating (\ref{eq:HTheur}) to $\bar{\gamma}$ it is straightforward to come up with the following relation
\beq
P_R= \frac{\bar{\gamma} \sigma^2}{1-  \frac{\gammabar}{1+\gammabar}\alpha}\;  \quad \Rightarrow \quad
p_k =  \frac{1}{h_k^2} \frac{\bar{\gamma} \sigma^2}{1- \frac{\gammabar}{1+\gammabar}\alpha} \; ,
\label{eq:tsepc}
\eeq
wherein, we recall, $\alpha=K/N$, and the relation $\alpha< 1 + 1/ \bar{\gamma}$ must hold.
Eq. (\ref{eq:tsepc}), which descends from eq. (16) in \cite{tse}, gives a simple power control algorithm that permits setting the transmitted power for each user based on the knowledge of the channel gain for the user of interest only.
The above algorithm, however, does not take into account the situation in which, due to fading and path losses, some users end up transmitting at their maximum power without achieving the target SINR, and indeed our numerical results to be shown in the sequel will prove the inability of (\ref{eq:tsepc}) to predict with good accuracy the actual power profile for the active users.

In order to circumvent this drawback, we first recall that in \cite{shamai} (see also \cite{husheng}) the following result has been shown:

\noindent
{\bf Lemma:} {\em Denote by $F(\cdot)$ the cumulative distribution function (CDF) of the squared fading coefficients $h_i^2$, and  by $[h^2_{[1]}, \, h^2_{[2]}, \, \ldots, \, h^2_{[K]}]$ the vector of the users' squared fading coefficients sorted in non-increasing order. Then we have that $h^2_{[\ell]}$ converges in probability (as $K \rightarrow \infty$)  to $F^{-1}\left(\frac{K-\ell}{K}\right)$, $\forall \ell=1, \ldots, K \, $.
}
\medskip

The above lemma states that if we sort a large number of identically distributed random variates, we obtain a vector that is approximately equal to the uniformly sampled version of the inverse of the common CDF of the random variates. Accordingly, in a large CDMA system each user may individually build a rough estimate of the fading coefficients in the network and thus will be able to predict the number of users, say $u_2$,  that possibly will wind up transmitting at the maximum power. Indeed, since, according to (\ref{eq:tsepc}) each user is to be received with a power $P_R$, the estimate $u_2$ of the number of users transmitting at the maximum power is given by
\beq
u_2= \ds \sum_{i=1}^{K}u \left( \frac{\bar{\gamma} \sigma^2}{F^{-1}\left(\frac{K-i}{K}\right) \left(1- \alpha
\frac{\gammabar}{1+\gammabar}\right)} - P_{\max} \right) \; ,
\label{eq:u2}
\eeq
with $u(\cdot)$ the step-function. It is also natural to assume that the users transmitting at $P_{\max}$ will be the ones with the smallest channel coefficients, i.e. the squared channel gains of the users transmitting at the maximum power are well approximated by the samples $F^{-1}\left(\frac{K-\ell}{K}\right)$, with $\ell=K-u_2+1, \ldots, K$. As a consequence, the generic $k$-th user will be affected by $u_1=K-u_2$ users that are received with power $P_R$ (these are the $u_1$ users with the strongest channel gains and that are able to achieve the target SINR $\gammabar$), and by $u_2$ users that are received with power $P_{\max} F^{-1}\left(\frac{K-\ell}{K}\right)$, with $\ell=K-u_2+1, \ldots, K$.
Denoting by $P_k$ the received power for the $k$-th user,
Eq. (\ref{eq:HTheur}) can be now written as
\beq
\gamma_k=\ds \frac{P_k}{
 \sigma^2 +\frac{u_1}{N} \frac{P_kP_R}{P_k + P_R\gamma_k}+ \frac{1}{N}{\ds \sum_{i=K-u_2+1}^K}
\frac{P_k P_{\max} F^{-1}\left(\frac{K-i}{K}\right)}{P_k+ P_{\max} F^{-1}\left(\frac{K-i}{K}\right)\gamma_k}
}\; .
\label{eq:casino}
\eeq
Now, assuming for the moment that user $k$ is able to achieve its target SINR, i.e. that $P_k=P_R$,
we can make the approximation
$\frac{P_kP_R}{P_k + P_R\gamma_k}\approx \frac{P_k }{1+ \gamma_k} \; ,
$
whereby equating (\ref{eq:casino})  to the target SINR $\gammabar$ we have
\beq
\ds \frac{ P_k}{
 \sigma^2 +\frac{u_1}{N} \frac{ P_k}{1 + \gammabar}+ \frac{1}{N}\sum_{i=K-u_2+1}^K
\frac{P_k P_{\max} F^{-1}\left(\frac{K-i}{K}\right)}{P_k+ P_{\max} F^{-1}\left(\frac{K-i}{K}\right)\gammabar}
} = \gammabar\; .
\label{eq:casino2}
\eeq
The above relation can now be solved numerically in order to determine the receive power $P_k$ for the  $k$-th user\footnote{Actually this equation gives the desired receive power for each user, and thus in a centralized power control algorithm needs to be solved only once.}. Finally, the actual transmit power for the $k$-th user is set according to the rule
\beq
p_k= \min \left\{P_k/h_k^2 , \, P_{\max} \right\} \ .
\label{eq:pupdate}
\eeq

The proposed algorithm may be summarized as follows. First, the number of users transmitting at the maximum power is estimated according to (\ref{eq:u2}). Then, the desired receive power for each user is computed by solving (\ref{eq:casino2}). Finally, the transmit power for the $k$-th user is determined according to relation (\ref{eq:pupdate}). Note that this algorithm requires knowledge only of the channel gain for the user of interest.


\section{Power control for SIC/MMSE detection}

Let us now consider the case in which non-linear SIC/MMSE detection is used at the receiver, and let us thus assume that users are ordered according to a non-increasing power profile. In this case the following theorem can be proved.

\noindent
{\bf Theorem:} {\em Let $\gamma_k$ be the (random) SINR of the SIC/MMSE receiver for the $k$-th user; let $P_k$ be the received power for the $k$-th user, and assume that previously detected symbols have been perfectly cancelled. As $K, N \rightarrow + \infty$, with $K/N=\alpha$, $\gamma_k$ converges in probability to $\gamma_k^*$, the unique solution of the equation
\beq
\gamma_k^*= \ds \frac{P_k}{\sigma^2 + \alpha_k E_{P|P < P_k} \left[
\frac{P P_k}{P_k +P \gamma_k^*}
\right]} \; ,
\label{eq:HTsic}
\eeq
with $\alpha_k=(K-k+1)/N$ and $E_{P|P < P_k}\left[ \cdot\right]$ denoting expectation with respect to the empirical distribution of the received powers not larger than $P_k$.
}

\noindent
{\bf Proof:} The proof is omitted for the sake of brevity.

To corroborate the statement of the above theorem, in Fig. 1 we report the asymptotic SINR and 100 actual realizations of  the SINR corresponding to random realizations of the spreading codes and of the received powers, assumed to follow a Rayleigh distribution. A system with processing gain $N=256$ has been considered here, and it is seen that the random SINR realizations are spread around their asymptotic value.

Using the notation of Section II, a heuristic reformulation of the above theorem states that in a large system the $k$-th user's SINR $\gamma_k$ is deterministic and approximately satisfies the equation
\beq
\gamma_k \approx \ds \frac{h^2_k p_k}{\sigma^2 +  \frac{1}{N}
\sum_{\ell=k+1}^{K} \frac{h^2_{\ell} p_{\ell} h^2_k p_k}{h^2_k p_k + h^2_{\ell} p_{\ell} \gamma_k}} \; .
\eeq

Now, the above expression can be used to derive a simple and effective power control algorithm. Let us consider the $K$-th user first, i.e. the one with the smallest channel gain. The information symbol from this user will be the last one to be detected, thus implying that, under the assumption of error-free detection of previous bits, and denoting by $P_K$ the received power for this user, the  relation
$
P_K/\sigma^2=\gammabar
$
should hold.
As a consequence, the transmit power for the $K$-th user is set to $p_K = \min \{\gammabar \sigma^2/h_K^2, P_{\max} \}$.

Consider now user $K-1$; on denoting by $P_{K-1}$ its received power, we have that
\beq
\frac{P_{K-1}}{ \sigma^2 + \frac{1}{N} \frac{P_{K-1} p_K h_K^2}{P_{K-1} + p_K h_K^2\gammabar}}=\gammabar \; .
\label{eq:uk-1}
\eeq
Now, we may reasonably assume that  $h_K$ and $h_{K-1}$ are approximately equal (recall that these are the two smallest channel gains), and thus that $P_{K-1} \approx P_K$. Moreover, in a distributed approach, we can substitute $h_K$ with its estimate $\hat{h}_K$ given by the Lemma of the previous section. As a consequence, (\ref{eq:uk-1})  can be approximated as
\beq
\frac{P_{K-1}}{ \sigma^2 + \frac{1}{N} \frac{P_{K-1} p_K \hat{h}_K^2}{P_{K} + p_K \hat{h}_K^2\gammabar}}=\gammabar\; .
\eeq
After solving the above equation for $P_{K-1}$, we can set the transmit power of the $(K-1)$-th user according to $p_{K-1}=\min\{P_{K-1}/h^2_{K-1} , P_{\max}\}$.

In general, for the generic $k$-th user, we have that the following relation must hold
\beq
\ds \frac{P_k}{\sigma^2 +  \frac{1}{N} \sum_{\ell=k+1}^{K}
\frac{P_k h_{\ell}^2 p_{\ell}}{P_k +h_{\ell}^2 p_{\ell} \gammabar}}=\gammabar   \; .
\eeq
Since it is reasonable to assume that $P_k \approx P_{k+1}$, and replacing the channel gains with their estimates, the above equation can be re-written as
\beq
\ds \frac{P_k}{\sigma^2 +  \frac{1}{N} \sum_{\ell=k+1}^{K}
\frac{P_k \hat{h}_{\ell}^2 p_{\ell}}{P_{k+1} +\hat{h}_{\ell}^2 p_{\ell} \gammabar}}=\gammabar   \; .
\eeq
Solving with respect to $P_k$ we have
\beq
P_k= \gammabar \sigma^2 \ds \frac{1}{1-  \frac{\gammabar}{N }\sum_{\ell=k+1}^{K}
 \frac{\hat{h}_{\ell}^2 p_{\ell}}{P_{k+1} + \hat{h}_{\ell}^2 p_{\ell}\gammabar}
} \, ,
\label{eq:formula1}
\eeq
and the transmit power for the $k$-th user is
\beq
p_k= \min\{P_k/h_k^2 \, , P_{\max}\} \, .
\label{eq:formula2}
\eeq

In summary, the proposed algorithm proceeds as follows. For centralized implementation, equations (\ref{eq:formula1}) and (\ref{eq:formula2}) are sequentially implemented for $k=K, K-1, \ldots, 2, 1$. Alternatively, for decentralized implementation, the generic $j$-th user must compute equations (\ref{eq:formula1}) and the equation
\beq
p_k= \min\{P_k/\hat{h}_k^2 \, , P_{\max}\} \, ,
\eeq
for $k=K, K-1, \ldots, j+1$.  Finally, equations (\ref{eq:formula1}) and (\ref{eq:formula2}) are implemented for $k=j$. Note that in the distributed implementation each user needs to know its own channel gain and also its order in the data detection sequence at the receiver, i.e. it must know how many  users interfere with it at the receiver. However, no information on its uplink SINR or on the parameters of the multiaccess interference is needed.

\begin{figure}[!tb]
\centerline{\hbox{\includegraphics[height=4cm,width=6cm]{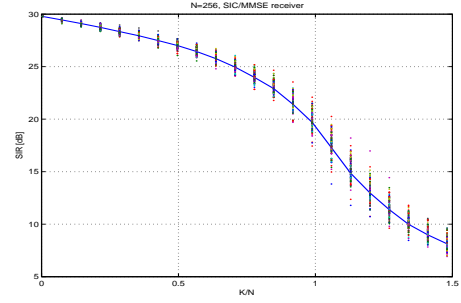}}}
\caption{Randomly generated SIC/MMSE SINR compared to the asymptotic limit, for the case of exponentially distributed powers (Rayleigh fading).}
\end{figure}

\begin{figure}[!tb]
\centerline{\hbox{\includegraphics[height=4.2cm,width=6cm]{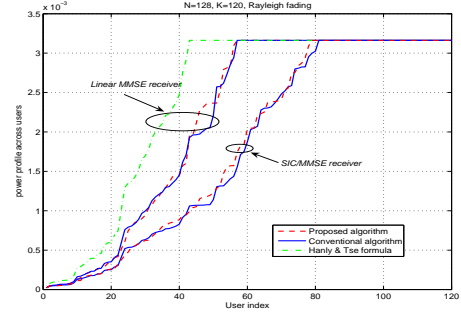}}}
\caption{Transmitted power profile across users for the proposed algorithm, the conventional power control algorithm \cite{yates} and the profile derived by the algorithm in \cite{tse}, for both cases of linear MMSE detection and SIC/MMSE detection.} \label{fig:9}
\end{figure}

\section{Numerical Results}
We consider an uplink DS/CDMA system with processing gain $N=128$, and assume that the packet length is $M=120$, and that the target SINR is
$\bar{\gamma}=6.689 = $8.25 dB. According to \cite{meshkati}, it can be seen that achieving the target SINR $\bar{\gamma}$ in the
considered scenario leads to a maximization of the utility, i.e. of the ratio between the packet success rate and the transmit power.

A single-cell system is considered, wherein users may have random positions with a distance from the BS ranging from 10m to 1000m. The channel coefficient $h_k$ for the generic $k$-th user is assumed to be Rayleigh distributed with mean equal to $d_k^{-1}$, with $d_k$ being the distance of user $k$ from the BS.
As to the thermal noise level, we take $\sigma^2=2 \cdot 10^{-9}$W/Hz, while the maximum allowed power $P_{k,\max}$ is $-25$dBW.
Fig. 2 reports the transmitted power profile across users for the algorithm proposed in Section III and IV, for the algorithm derived by Eq. (16) in \cite{tse} (i.e. eq. (\ref{eq:tsepc})), and for a conventional algorithm (see \cite{yates}) that is non-adaptive and requires a substantial amount of prior information. It is seen that the proposed algorithms are capable of reproducing the optimal power profile with very good accuracy, while, on the contrary, the algorithm descending from paper \cite{tse} overestimates the required transmit powers and does not exhibit good performance.
It is seen clearly that adopting a SIC/MMSE receiver yields considerable savings in transmit power needed to achieve a certain target SINR.

While Fig. 2  shows the result of only one simulation trial (note however that  similar behavior has been observed in every case we considered), the remaining three figures report results coming from an average over 1000 independent trials. Figs. 3 - 5 show the achieved average utility (measured in bits/Joule), the average per-user transmit power and the average achieved SINR at the receiver output for the conventional power control algorithms (for both linear MMSE and SIC/MMSE detection) \cite{yates}, for the proposed algorithms (for both linear MMSE and SIC/MMSE detection), and for the power control algorithm derived by paper \cite{tse}. These results show that the proposed algorithms achieve performance levels practically indistinguishable from those of the standard algorithms, while the algorithm (\ref{eq:tsepc}) achieves a much smaller  utility.
From Fig. 5 it is however seen that the algorithm  (\ref{eq:tsepc}) achieves an output SINR larger than that of any other algorithm considered: this should not be interpreted as a sign of good performance. Indeed, in the considered scenario the aim of the power control algorithm is to make each user operate at a SINR equal to $\gammabar$.


\section{Conclusions}
This paper has considered the design of distributed power control algorithms for cellular CDMA systems based on asymptotic analysis, for  situations in which either linear MMSE detection or a non-linear SIC/MMSE detection are used by the receiver. For the latter case, closed-form formulas for the limiting system SINR have also been developed.



Overall, the proposed solutions achieve satisfactory performance, and the proposed approach is quite promising.  Among the authors' current research efforts in this area is the extension of the proposed algorithms to the situation in which the received signals have been affected by multipath.

\section*{Acknowledgements}
This research was supported in part by the U. S. Air Force Research Laboratory under Cooperative Agreement No.
FA8750-06-1-0252 and in part by the U. S. Defense Advanced Research Projects Agency under Grant HR0011-06-1-0052.

\bibliographystyle{IEEEtran}

\begin{figure}[!tb]
\centerline{\hbox{\includegraphics[height=4cm,width=6cm]{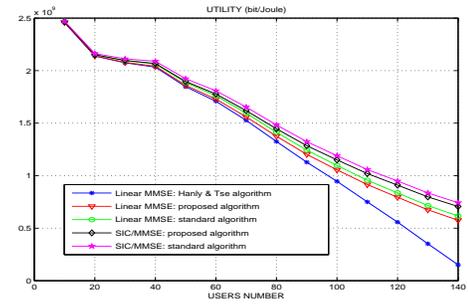}}}
\caption{Average utility versus number of users.}
\end{figure}

\begin{figure}[!tb]
\centerline{\hbox{\includegraphics[height=4cm,width=6cm]{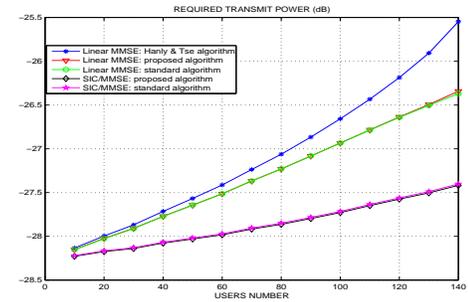}}}
\caption{Average transmit power versus number of users.}
\end{figure}

\begin{figure}[!tb]
\centerline{\hbox{\includegraphics[height=4cm,width=6cm]{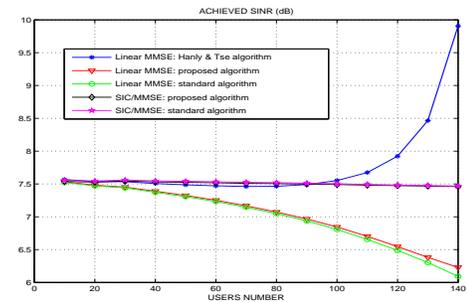}}}
\caption{Average achieved SINR versus number of users.}
\end{figure}

\end{document}